\documentclass[preprint,eqsecnum,aps,nofootinbib]{revtex4}
\usepackage{amsfonts,amsmath,amssymb,amsthm}
\usepackage{latexsym}
\usepackage{bbm,bm}
\usepackage{graphicx}

\newcommand{\ket}[1]{\lvert #1 \rangle}

\newcommand{\beq}{\begin{equation}}
\newcommand{\eeq}{\end{equation}}
\newcommand{\beqs}{\begin{eqnarray}}
\newcommand{\eeqs}{\end{eqnarray}}

\begin{document}

\title{Asymmetric Quantum Illumination with three-mode Gaussian State}

\author{Mi-Ra Hwang$^1$ and DaeKil Park$^{1,2}$\footnote{corresponding author, dkpark@kyungnam.ac.kr} }

\affiliation{$^1$Department of Electronic Engineering, Kyungnam University, Changwon,
                 631-701, Korea    \\
                 $^2$Department of Physics, Kyungnam University, Changwon,
                  631-701, Korea }

\begin{abstract}
Quantum illumination with asymmetric strategy  is examined by making use of a three-mode maximally entangled Gaussian state, which involves one signal and two idler beams. 
It is shown that this scenario gives less-error probability compared to that with  a two-mode squeezed vacuum state 
when $N_S$, average photon number per signal, is less than $0.46$.

\end{abstract}
\maketitle

\section{Introduction}
Quantum entanglement\cite{schrodinger-35,text,horodecki09} is at the heart in the various types of quantum information processing (QIP) and recently developed quantum technologies (QT).
It is used as a physical resource in  many QIP and QT such as  in quantum teleportation\cite{teleportation,Luo2019},
superdense coding\cite{superdense}, quantum cloning\cite{clon}, quantum cryptography\cite{cryptography,cryptography2}, quantum
metrology\cite{metro17}, and quantum computers\cite{qcreview,computer,supremacy-1}. In particular, quantum computing attracted a lot of  attention recently after IBM and Google independently realized gate-based quantum computers.
It is debatable whether  ``quantum supremacy'' is achieved or not in the quantum computation. 

A few years ago another type of  entanglement-assisted QIP called quantum illumination (QI)\cite{lloyd08} became of interest to the research community. 
The purpose of this protocol is to detect  low reflective objects embedded in baths of  strong thermal noise.
Therefore, the protocol can be described by using a hypothesis testing\cite{hypo1,hypo2}. Let the null hypothesis $H_0$ be target-absence and the alternative hypothesis $H_1$ be target-presence.
Then, there are two type of errors. First one is type-I error (or false alarm), whose probability is expressed as a conditional probability $P(H_1 | H_0)$. 
Second one is type-II error (or false negative), whose probability is $P(H_0 | H_1)$. In this reason there exist two strategies in the detection process. First one called symmetric strategy\cite{tan08} is to minimize the average error-probability
$P_E = P(H_0) P(H_1 | H_0) + P(H_1) P(H_0 | H_1)$, where $P(H_0)$ and $P(H_1)$ are the prior probabilities associated with the two hypotheses. Second one called asymmetric strategy\cite{asymmetry-1,asymmetry-2} is to minimize the type-II error-probability with 
accepting the type-I error-probability to some extent. In the following we will call the former by symmetric QI (SQI) and the latter by asymmetric QI (AQI).

In Ref.\cite{tan08} the SQI was examined by using  a two-mode squeezed vacuum (TMSV) state 
\begin{equation}
\label{two-signal-state}
\ket{\psi}_{SI} = \sum_{n=0}^{\infty} \sqrt{\frac{N_S^n}{(1 + N_S)^{n+1}}} \ket{n}_S \ket{n}_I.
\end{equation}
This is two-mode zero-mean maximally entangled Gaussian state\cite{gaussian1}. In Eq. (\ref{two-signal-state}) the subscripts $S$ and $I$ stand for signal and idler modes respectively.
Also $N_S$ represents the average photon number per signal mode.
The typical SQI can be described as follows. The transmitter generates two entangled photons called signal (S) and idler (I) modes. The S-mode photon is used to interrogate 
the unknown object hidden in the background. After receiving photon from a target region, joint quantum measurement for returned light and retained I-mode photon is performed to decide absence or presence of target.
The most surprising result of the SQI is the fact that the error-probability $P_E$ for the detection is drastically lowed even if the initial entanglement between S and I modes disappears 
due to strong decoherence. It was shown in Ref.\cite{tan08} that compared to the classical coherent-state illumination  the SQI with the TMSV state achieves $10 \log_{10} 4 \approx 6.02$ dB
quantum advantage in term of the error-probability if $N_B \gg N_S$, where $N_B$ is the average photon number of background thermal state. 
An experimental realization of QI was also explored in Ref.\cite{guha09,lopaeva13,barzanjeh15,zhang15,zhuang17}.

The AQI with the TMSV state was also studied in Ref.\cite{asymmetry-1,asymmetry-2}. In this case the error-probability $P_{err}$ is estimated from quantum Stein's lemma\cite{li12,TH12,CMMAB08,JOPS12,DPR15} as follows. 
Let $\rho$ and $\sigma$ be quantum states for target absence and target presence respectively. If there are $M$ identical copies of $\rho$ and $\sigma$, and $\varepsilon$ is a permitted range of the type-I error-probability, the error-probability is 
\begin{equation}
\label{asymmetric_error}
P_{err} = e^{-M R}
\end{equation}
where
\begin{equation}
\label{error_1}
R = a + \sqrt{ b / M} \Phi^{-1} (\varepsilon) + {\cal O} (M^{-1} \ln M).
\end{equation}
In Eq. (\ref{error_1}) $\Phi (y)$ is the cumulative distribution function for a standard normal random variable $\Phi (y) \equiv (1 / \sqrt{2 \pi}) \int_{-\infty}^y dx \exp(-x^2 / 2)$, and $a$ and $b$ are
the quantum relativity entropy and quantum relative entropy variance defined as 
\begin{eqnarray}
\label{relative-1}
&& a = D(\rho || \sigma) = \mbox{Tr} [\rho \ln \rho - \rho \ln \sigma]            \\   \nonumber
&&b = V (\rho || \sigma) = \mbox{Tr} \left[\rho \left\{\ln \rho - \ln \sigma - D(\rho || \sigma) \right\}^2 \right].
\end{eqnarray}
As Ref.\cite{asymmetry-1,asymmetry-2} have shown,  compared to the classical coherent-state illumination $P_{err}$ is reduced if the TMSV state is used as an initial state. This means that there exist an quantum advantage in this case too even though $\rho$ and $\sigma$
loose their entanglement due to strong entanglement-breaking noise. The main results of Ref.\cite{asymmetry-1,asymmetry-2} will be briefly reviewed in next section as a figure 1. 

There are a lot of issues in the QI process we need to discuss such as non-Gaussian state approach\cite{zhang14,fan18}, microwave illumination\cite{micro}, and quantum receivers\cite{guha09,zhuang17,guha09-2,Jo21-2} where the joint quantum measurement is performed.
Among them the most important issue from an aspect of quantum information theories is, in our opinion, the following question: what is the physical resource in the illumination process? Recently, it was shown that entanglement is not unique resource\cite{museong23-1} in the QI by making use of the 
various squeezing operators. Authors in Ref.\cite{discord1,discord2} suggested that quantum discord is a genuine resource responsible for the quantum advantage. 
However, it was argued in Ref.\cite{anti-discord1} that the advantage cannot be characterized by a quantum discord sorely. Furthermore, the counterexample was found in Ref.\cite{Jo21-1}, which supports Ref.\cite{anti-discord1}.
 It was shown in Ref.\cite{palma18,nair20,brad21} that the TMSV state is a nearly optimal state in the error-probability provided that reflectivity is extremely small. 
 
\begin{figure}[ht!]
\begin{center}
\includegraphics[height=5.0cm]{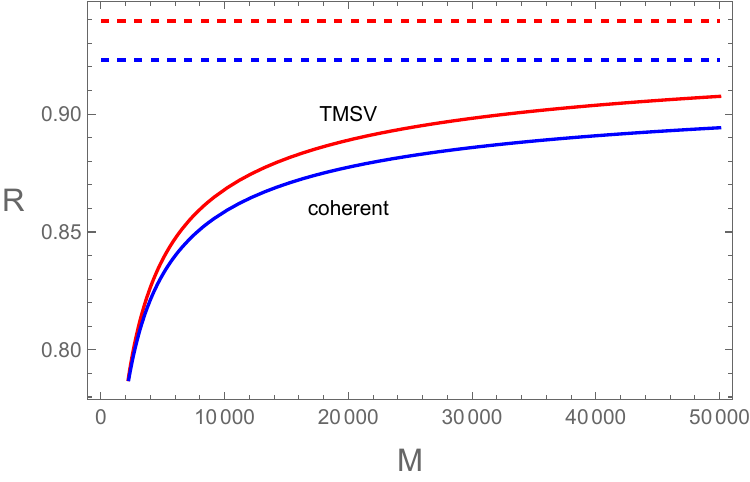}  \hspace{.2cm}
\includegraphics[height=5.0cm]{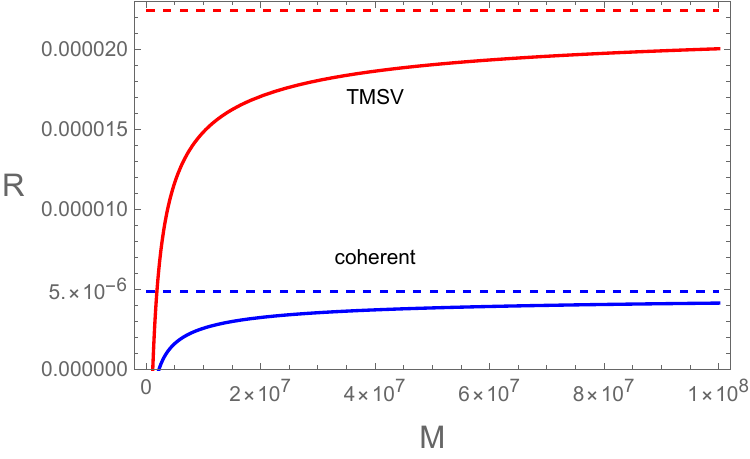}
\caption[fig1]{(Color online) (a) $M$-dependence of $R$ when $N_B = 0.01$, $N_S = 20$, $\kappa = 0.01$, and $\varepsilon = 0.001$. The red and blue lines 
correspond to TMSV and classical coherent states. The red dashed and blue dashed lines correspond to the maximum values of $R$ for TMSV and classical coherent states, which are $0.9395$ and $0.9230$ respectively.        
(b)  $M$-dependence of $R$ when $N_B = 20$, $N_S = 0.01$, $\kappa = 0.01$, and $\varepsilon = 0.01$. The red and blue lines 
correspond to TMSV and classical coherent states. The red dashed and blue dashed lines correspond to the maximum values of $R$ for TMSV and coherent states, which are $2.24 \times 10^{-5}$ and $4.88 \times 10^{-6}$ respectively.           }
\end{center}
\end{figure}
 
However, as shown in Ref.\cite{eylee21} SQI with three-mode maximally entangled Gaussian state
 gives less error-probability compared to that with TMSV state when $N_S \ll N_B$ and $N_S < 0.295$.
 We conjecture that  the three-mode Gaussian state may also give an enhancement of quantum advantage compared to the TMSV state in the AQI. 
 We will show in this paper that this is indeed the case when  $N_S \ll N_B$ and $N_S < 0.46$. In section II we introduce two main results of Ref.\cite{asymmetry-1,asymmetry-2} in Fig. 1. In section III we compute the error-probability of the AQI when 
 the three-mode maximally entangled Gaussian state is used as an initial state. It is shown in this section that the error-probability for this case is less than that for the TMSV state when  $N_S \ll N_B$ and $N_S < 0.46$. In section IV a brief conclusion is given.

\section{AQI with TMSV state}
The AQI with the TMSV state was explored in Ref.\cite{asymmetry-1,asymmetry-2}. We do not want to repeat the calculation here. 
What we want is to introduce the major results of Ref.\cite{asymmetry-1,asymmetry-2}, which is summarized in Fig. 1.

In Fig. 1(a) we plot the $M$-dependence of $R$ introduced in Eq. (\ref{error_1}) when $N_B = 0.01$, $N_S = 20$, $\kappa = 0.01$, and $\varepsilon = 0.001$, where $\kappa$ is a reflectivity from a target and $\varepsilon$ is a permitted range of the  the type-I error-probability.
The red and blue lines correspond to $R$ of TMSV and classical coherent states. Since the red line is larger than the blue lines in the entire range of $M$, this fact indicates that the error-probability given in Eq. (\ref{asymmetric_error}) becomes small in the TMSV state compared to the classical coherent state. 
The red dashed and blue dashed lines correspond to the maximum values of $R$, which are the quantum relative entropy $D(\rho || \sigma)$ for each states. In Fig. 1(a)  the maximum values of $R$ are $0.9395$ and $0.9230$ respectively. 
Fig. 1(b) is a plot of $R$ with respect to $M$ when $N_B = 20$, $N_S = 0.01$, $\kappa = 0.01$, and $\varepsilon = 0.01$. In this case also $R$ for the TMSV state is much larger than $R$ for the classical coherent state, which implies that there exist a quantum advantage in error-probability. 
The maximal values of $R$ become $2.24 \times 10^{-5}$ and $4.88 \times 10^{-6}$ respectively. 

In next section we will introduce the three-mode maximally entangled Gaussian state and compute the corresponding $R$. We will compare it to the results of TMSV state.

\section{AQI with three-mode maximally entangled Gaussian state}
The three-mode maximally entangled Gaussian state $\rho_G$ was introduced in Ref.\cite{eylee21} to explore the SQI. 
It is a zero-mean Gaussian state whose covariance matrix takes the following block form\footnote{Let $\hat{x}_j$ denote each quadrature operator and the vector of the quadrature operator is chosen as 
$\hat{x} = [\hat{q}_1, \hat{q}_2, \hat{q}_3, \hat{p}_1, \hat{p}_2, \hat{p}_3] \equiv [\hat{x}_1, \cdots, \hat{x}_6]$, where $\hat{q}$ and $\hat{p}$ are position and momentum operators.    }:
\begin{eqnarray}
\label{signal-state}
\Lambda_{S I_1 I_2}^{(3)} = \left( \begin{array}{ccc}
                          S  &   C  &  C         \\
                          C  &  S  &   C        \\
                          C  &  C  &  S         
                                                \end{array}                \right) \oplus
                                          \left( \begin{array}{ccc}
                          S  &  - C  &  -C         \\
                          -C  &  S  &  - C        \\
                          -C  & - C  &  S         
                                                \end{array}                \right)
\end{eqnarray}
where $A \oplus B = \left( \begin{array}{cc} A & 0 \\ 0 & B \end{array} \right)$ and $S = N_S + \frac{1}{2}$. 
As commented in Ref.\cite{eylee21} the off-diagonal term $C$ is restricted by $0 \leq C \leq C_{max}$, where $C_{max}^2$ is a root of the cubic equation $4x^3 - 9 S^2 x^2 + 6 S^4 x - (S^6 - 1/64) = 0$\footnote{In Ref.\cite{eylee21} $\widetilde{x} = \left( \widetilde{C}_{max} \right)^2$ satisfies 
$4 \widetilde{x}^3 - 9 \widetilde{S}^2 \widetilde{x}^2 + 6 \widetilde{S}^4 \widetilde{x} - (\widetilde{S}^6 - 1) = 0$, where $\widetilde{S} \equiv 1 + 2 N_S = 2 S$. This condition is derived from the fact that the Gaussian state with covariance matrix $2 \Lambda_{S I_1 I_2}^{(3)}$ is pure state. 
Using $\widetilde{x} = 4 x$ and $\widetilde{S} = 2 S$, one can derive $4 x^3 - 9 S^2 x^2 + 6 S^4 x - (S^6 - 1 / 64) = 0$, where $x = \left( C_{max} \right)^2$.} .
The explicit expression of $C_{max}$ is 
\begin{equation}
\label{ex_cmax}
C_{max}^2 = \frac{1}{4} \left[ 3 S^2 - 4 S^4 \eta^{-2/3} - \frac{1}{4} \eta^{2/3} \right]
\end{equation}
where $\eta = 2 \left[ \sqrt{1 + 16 S^6} - 1 \right]$.
In order to examine the entanglement of $\rho_G$ one can compute the logarithmic negativity\cite{gaussian1} explicitly, which is one of the entanglement measure. From the  logarithmic negativity  one can show that $\rho_G$ is separable when $0 \leq C \leq C_c$, where
\begin{equation}
\label{boundary}
C_c^2 = \frac{1}{8} \left[ (2 + 5 N_S + 5 N_S^2) - \sqrt{(1 + 3 N_S) (2 + 3 N_S) (2 + N_S + N_S^2)} \right].
\end{equation}
In the region $C_c < C \leq C_{max}$ $\rho_G$ is entangled state and it is maximized at $C = C_{max}$. In the following we will fix $C$ as $C = C_{max}$.

Let $\rho$ and $\sigma$ be quantum states for target absence and target presence respectively. Both are the zero-mean Gaussian states. Since, for $\rho$, the annihilation operator 
for the return from the target region should be $\hat{a}_R = \hat{a}_B$, where $\hat{a}_B$ is the annihilation operator for a thermal state with average photon number $N_B$, its covariance matrix 
can be written in a form:
\begin{eqnarray}
\label{rho-state}
V_{\rho}  = \left( \begin{array}{ccc}
                          B  &   0  &  0         \\
                          0  &  S  &   C        \\
                          0  &  C  &  S         
                                                \end{array}                \right) \oplus
                                          \left( \begin{array}{ccc}
                          B  &  0  &  0         \\
                          0  &  S  &  - C        \\
                          0  & - C  &  S         
                                                \end{array}                \right)
\end{eqnarray}
where $B = N_B + \frac{1}{2}$.

For $\sigma$ the return-mode's annihilation operator would be $\hat{a}_R = \sqrt{\kappa} \hat{a}_S + \sqrt{1 - \kappa} \hat{a}_B$, 
where $\hat{a}_B$ is an annihilation operator for a thermal state with average photon number $N_B / (1 - \kappa)$. We assume $\kappa \ll 1$.
Combining all of the facts, one can deduce that the covariance matrix for $\sigma$ is 
\begin{eqnarray}
\label{sigma-state}
V_{\sigma}  = \left( \begin{array}{ccc}
                          A  &   \sqrt{\kappa} C  &  \sqrt{\kappa} C         \\
                          \sqrt{\kappa} C  &  S  &   C        \\
                          \sqrt{\kappa} C  &  C  &  S         
                                                \end{array}                \right) \oplus
                                          \left( \begin{array}{ccc}
                          A  &  -\sqrt{\kappa} C  &  -\sqrt{\kappa} C         \\
                          -\sqrt{\kappa} C  &  S  &  - C        \\
                         - \sqrt{\kappa} C  & - C  &  S         
                                                \end{array}                \right)
\end{eqnarray}
where $\kappa$ is reflectivity from a target and $A = \kappa N_S + B$.

The normal mode decomposition $V_{\rho} = S_{\rho} \left(D_{\rho} \oplus D_{\rho} \right) S_{\rho}^T$ can be derived, where
$S_{\rho}$ ($\in Sp_{6, \mathbb{R}}$, i.e. $S_{\rho} \Omega S_{\rho}^T = \Omega$ ) is given by 
\begin{eqnarray}
\label{sympletic_rho1}
S_{\rho} = \left( \begin{array}{ccc}
                          1  &   0  &  0         \\
                          0  &  \frac{1}{\sqrt{2} z}  &   -\frac{z}{\sqrt{2}}        \\
                          0  &  \frac{1}{\sqrt{2} z}  &    \frac{z}{\sqrt{2}}       
                                                \end{array}                \right) \oplus
                                          \left( \begin{array}{ccc}
                          1  &  0  &  0         \\
                          0  &  \frac{z}{\sqrt{2}}  &  - \frac{1}{\sqrt{2} z}        \\
                          0  & \frac{z}{\sqrt{2}}  &  \frac{1}{\sqrt{2} z}         
                                                \end{array}                \right)
\end{eqnarray}
with 
\begin{equation}
\label{def_z-1}
z = \left( \frac{S - C}{S + C} \right)^{1/4}
\end{equation}
and $D_{\rho} = \mbox{diag} \left(B, \sqrt{S^2 - C^2}, \sqrt{S^2 - C^2}\right)$.
Then, one can compute the followings straightforwardly:
\begin{eqnarray}
\label{rho_remainings1}
&&Z_{\rho} \equiv \det\left( V_{\rho} + \frac{i}{2} \Omega \right) = \frac{1}{64} (4 B^2 - 1) (4 S^2 - 4 C^2 - 1)^2    \\   \nonumber
&&G_{\rho} \equiv - 2 \Omega S_{\rho} \left[ \coth^{-1} (2 D_{\rho}) \right]^{\oplus 2} S_{\rho}^T \Omega         
 = \left( \begin{array}{ccc}
                          \rho_1  &   0  &  0         \\
                          0  &  \rho_2  &   -\rho_3        \\
                          0  &  -\rho_3  &  \rho_2         
                                                \end{array}                \right) \oplus
                                          \left( \begin{array}{ccc}
                          \rho_1  &  0  &  0         \\
                          0  &  \rho_2  &  \rho_3        \\
                          0  & \rho_3  &  \rho_2         
                                                \end{array}                \right)
\end{eqnarray}
where $\Omega = \left[ \begin{array}{cc} 0 & 1 \\ -1 & 0 \end{array} \right] \otimes I_3$, $\coth^{-1} x \equiv \frac{1}{2} \ln \left( \frac{x+1}{x-1}\right)$ and
\begin{eqnarray}
\label{rho_boso1}
&& \rho_1 = \ln \frac{N_B + 1}{N_B}                                    \\    \nonumber
&& \rho_2 = \frac{S}{\sqrt{S^2 - C^2}} \ln \frac{2 \sqrt{S^2 - C^2} + 1}{2 \sqrt{S^2 - C^2}  - 1}     \\   \nonumber
&& \rho_3 =  \frac{C}{\sqrt{S^2 - C^2}} \ln \frac{2 \sqrt{S^2 - C^2} + 1}{2 \sqrt{S^2 - C^2}  - 1}.
\end{eqnarray}

\begin{figure}[ht!]
\begin{center}
\includegraphics[height=5.0cm]{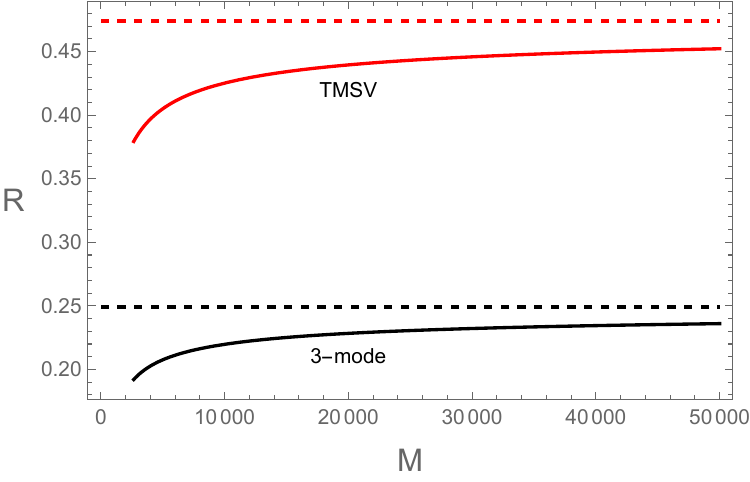}   \hspace{.2cm}
\includegraphics[height=5.0cm]{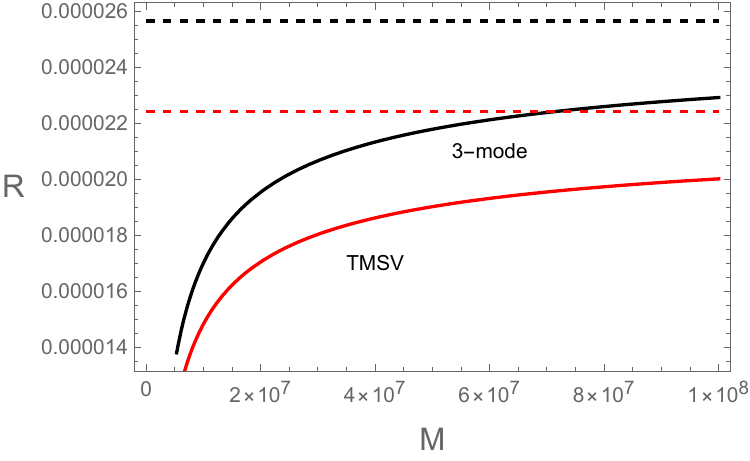}
\caption[fig2]{(Color online) (a) $M$-dependence of $R$ when $N_B = 0.01$, $N_S = 10$, $\kappa = 0.01$, and $\varepsilon = 0.001$. The red and black lines 
correspond to TMSV and $3$-mode Gaussian state. The red dashed and black dashed lines correspond to the maximum values of $R$ for TMSV and $3$-mode Gaussian state, which are $0.474$ and $0.249$ respectively.        
(b)  $M$-dependence of $R$ when $N_B = 20$, $N_S = 0.01$, $\kappa = 0.01$, and $\varepsilon = 0.01$. The red and black lines 
correspond to TMSV and $3$-mode Gaussian state. The red dashed and black dashed lines correspond to the maximum values of $R$ for TMSV and $3$-mode Gaussian, which are $2.24 \times 10^{-5}$ and $2.57 \times 10^{-5}$ respectively.           }
\end{center}
\end{figure}

For $V_{\sigma}$ the normal mode decomposition $V_{\sigma} = S_{\sigma} \left(D_{\sigma} \oplus D_{\sigma} \right) S_{\sigma}^T$ can be derived, where
\begin{eqnarray}
\label{sympletic_sigma1}
&&S_{\sigma} = \left( \begin{array}{ccc}
                          0  &   x_+  &  y_-         \\
                          \frac{z}{\sqrt{2}}  &  u_+  &   v_-        \\
                          -\frac{z}{\sqrt{2}}  &  u_+  &    v_-       
                                                \end{array}                \right) \oplus
                                          \left( \begin{array}{ccc}
                          0  &  y_+  &  x_-         \\
                          \frac{1}{\sqrt{2} z}  &  v_+  &  u_-        \\
                          -\frac{1}{\sqrt{2} z}  & v_+  &  u_-         
                                                \end{array}                \right)                          \\   \nonumber
&&D_{\sigma} = \mbox{diag} (\beta_1, \beta_2, \beta_3).
\end{eqnarray}
In Eq. (\ref{sympletic_sigma1}) $z$ is given in Eq. (\ref{def_z-1}) and remaining parameters are 
\begin{eqnarray}
\label{sigma-remaining1}
&& x_{\pm} = \pm \frac{1}{2} \sqrt{\frac{\mu_{1,\pm} \mu_{2,\pm}}{(A - S \mp C) \xi \beta_{\pm}}}
\hspace{1.0cm} y_{\pm} = \sqrt{\frac{(A - S \mp C) \mu_{2, \pm} \beta_{\pm}}{\mu_{1,\pm} \xi}}                \\   \nonumber
&&u_{\pm} = \sqrt{\frac{\mu_{1,\pm} \mu_{2,\mp}}{8 (A - S \pm C) \xi \beta_{\pm}}}
\hspace{1.0cm} v_{\pm} = \mp \sqrt{\frac{(A - S \pm C) \mu_{2,\mp} \beta_{\pm}}{2 \mu_{1,\pm} \xi}}    \\   \nonumber
&&\beta_1 = \sqrt{S^2 - C^2}  \hspace{1.0cm} \beta_2\equiv \beta_+ = \sqrt{\frac{A^2 + S^2 - (1 + 4 \kappa) C^2 + \xi}{2}}  \\   \nonumber 
&&\hspace{1.0cm}\beta_3\equiv \beta_- = \sqrt{\frac{A^2 + S^2 - (1 + 4 \kappa) C^2 - \xi}{2}}
\end{eqnarray}
where
\begin{eqnarray}
\label{sigma-remaining2}
&&\mu_{1,\pm} =  (\xi - 2 A C) \pm [(A - S)^2 - C^2]  \hspace{1.0cm} \mu_{2,\pm} = A^2 - S^2 + C^2 \pm \xi     \\    \nonumber
&&\hspace{1.0cm} \xi = \sqrt{(A^2 - S^2 + C^2)^2 - 8 \kappa C^2 (A - S + C) (A - S - C)}.
\end{eqnarray}
The following useful formulas can be directly derived: 
\begin{eqnarray}
\label{useful-1}
&&   \mu_{2,+} - \mu_{2,-} = 2 \xi \hspace{1.0cm} \mu_{2,+} \mu_{2,-} = 8 \kappa C^2 (A - S + C) (A - S - C)            \\   \nonumber
&& \mu_{1,+} \mu_{2,+} = 2 (A - S - C) \left[ A \mu_{2,+} - 4 \kappa C^2 (A - S + C) \right]                         \\  \nonumber
&&\mu_{1,-} \mu_{2,-} = -2 (A - S + C) \left[ A \mu_{2,-} - 4 \kappa C^2 (A - S - C) \right].
\end{eqnarray}
Then,  one can show straightforwardly
\begin{eqnarray}
\label{sigma-remainings3}
&&Z_{\sigma} \equiv \det\left( V_{\sigma} + \frac{i}{2} \Omega \right) \\   \nonumber
&&= \frac{1}{64}  (4 S^2 - 4 C^2 - 1) \left[(4 A^2 - 1) (4 S^2 - 4 C^2 - 1) - 16 \kappa C^2 (4 A S - 1) + 64 \kappa^2 C^4 \right]   \\   \nonumber
&&G_{\sigma} \equiv - 2 \Omega S_{\sigma} \left[ \coth^{-1} (2 D_{\sigma}) \right]^{\oplus 2} S_{\sigma}^T \Omega         
 = \left( \begin{array}{ccc}
                          \sigma_1  &   \sigma_5  &  \sigma_5         \\
                          \sigma_5  &  \sigma_2  &   \sigma_6        \\
                          \sigma_5  &  \sigma_6  &  \sigma_2         
                                                \end{array}                \right) \oplus
                                          \left( \begin{array}{ccc}
                          \sigma_3  &  \sigma_7  &  \sigma_7         \\
                          \sigma_7  &  \sigma_4  &  \sigma_8        \\
                          \sigma_7  & \sigma_8  &  \sigma_4         
                                                \end{array}                \right)
\end{eqnarray}
where
\begin{eqnarray}
\label{sigma-remaining4}
&& \sigma_1 = \gamma_2 y_+^2 + \gamma_3 x_-^2     \hspace{2.0cm}  \sigma_2 = \frac{\gamma_1}{2 z^2} + \gamma_2 v_+^2 + \gamma_3 u_-^2      \\   \nonumber
&&\sigma_3 = \gamma_2 x_+^2 + \gamma_3 y_-^2      \hspace{2.0cm} \sigma_4 = \frac{z^2 \gamma_1}{2} + \gamma_2 u_+^2 + \gamma_3 v_-^2       \\   \nonumber
&&\sigma_5 = \gamma_2 y_+ v_+ + \gamma_3 x_- u_-    \hspace{1.0cm}  \sigma_6 = - \frac{\gamma_1}{2 z^2} + \gamma_2 v_+^2 + \gamma_3 u_-^2   \\   \nonumber
&&\sigma_7 = \gamma_2 x_+ u_+ + \gamma_3 y_- v_-    \hspace{1.0cm} \sigma_8 = - \frac{z^2 \gamma_1}{2} + \gamma_2 u_+^2 + \gamma_3 v_-^2.
\end{eqnarray}
In Eq. (\ref{sigma-remaining4}) $\gamma_j$ arises from $\coth^{-1} (2 D_{\sigma}) = \frac{1}{2} \mbox{diag} (\gamma_1, \gamma_2, \gamma_3)$, where
\begin{equation}
\label{sigma-remaining5}
\gamma_j = \ln  \frac{2 \beta_j + 1}{2 \beta_j - 1}.          \hspace{.3cm} (j = 1, 2, 3)
\end{equation}

Now, we define $\Gamma = G_{\rho} - G_{\sigma}$. Then, it is straightforward to show 
\begin{eqnarray}
\label{relative-11}
&&\mbox{Tr} (\Gamma V_{\rho}) = B (2 \rho_1 - \sigma_1 - \sigma_3) + 2 \big[ S (2 \rho_2 - \sigma_2 - \sigma_4) - C (2 \rho_3 + \sigma_6 - \sigma_8) \big]    \\     \nonumber
&&\mbox{Tr} \left[ \left( \Gamma V_{\rho} \right)^2 \right] = B^2 \left[ (\rho_1 - \sigma_1)^2 + (\rho_1 - \sigma_3)^2 \right]       + 4 B \left[ (S + C) \sigma_5^2 + (S - C) \sigma_7^2 \right]    \\        \nonumber
&&\hspace{2.5cm} + 2 \left[ S (\rho_2 - \sigma_2) - C (\rho_3 + \sigma_6) \right]^2  + 2 \left[ S (\rho_3 + \sigma_6) - C (\rho_2 - \sigma_2) \right]^2                                                                 \\        \nonumber
&&\hspace{2.5cm} + 2 \left[ S (\rho_2 - \sigma_4) - C (\rho_3 - \sigma_8) \right]^2 + 2 \left[S (\rho_3 - \sigma_8) - C (\rho_2 - \sigma_4) \right]^2                                                                    \\       \nonumber
&&\mbox{Tr} \left[ \left( \Gamma \Omega \right)^2 \right] = 4 (\rho_3 + \sigma_6) (\rho_3 - \sigma_8) - 2 (\rho_1 - \sigma_1) (\rho_1 - \sigma_3) - 4 (\rho_2 - \sigma_2) (\rho_2 - \sigma_4) - 8 \sigma_5 \sigma_7.                                                                                                                     
\end{eqnarray}
It was shown in Ref.\cite{asymmetry-1,asymmetry-2} that if $\rho$ and $\sigma$ are Gaussian states with corresponding means  $(\mu_{\rho}, \mu_{\sigma})$ and covariance matrices $(V_{\rho}, V_{\sigma})$, then 
$D(\rho || \sigma)$ and $V(\rho || \sigma)$ become
\begin{eqnarray}
\label{relative-12}
&&D(\rho || \sigma) = \frac{1}{2} \left[ \ln \frac{Z_{\sigma}}{Z_{\rho}} - \mbox{Tr} \left\{\Gamma V_{\rho} \right\} + \gamma^T G_{\sigma} \gamma \right]              \\   \nonumber
&&V(\rho || \sigma) = \frac{\mbox{Tr} \left\{\left(\Gamma V_{\rho} \right)^2 \right\}}{2} + \frac{\mbox{Tr} \left\{ \left( \Gamma \Omega \right)^2 \right\}}{8} + \gamma^T G_{\sigma} V_{\rho} G_{\sigma} \gamma
\end{eqnarray}
where $\gamma \equiv \mu_{\rho} - \mu_{\sigma}$. Since $\rho$ and $\sigma$ for our case are zero-mean Gaussian states, $\gamma = 0$. 
Then, using Eq. (\ref{rho_remainings1}), Eq. (\ref{sigma-remainings3}), and Eq. (\ref{relative-11}), one can compute $a = D(\rho || \sigma)$ and $b = V(\rho || \sigma)$ explicitly.
In this way it is possible to compute the exponent of error-probability $R$ given in Eq. (\ref{error_1}) analytically. Due to the lack of space  the explicit expressions of $R$ will not be presented in this paper because it is too lengthy. 

The $M$-dependence of $R$ is plotted in Fig. 2. Fig. 2(a) shows that  $R$ for the three-mode Gaussian state is smaller than that for the two-mode TMSV state when $N_B = 0.01$, $N_S = 10$, $\kappa = 0.01$, and $\varepsilon = 0.001$. 
Fig. 2(b) exhibits that  $R$ for the three-mode Gaussian state is larger than that for the two-mode TMSV state when $N_B = 20$, $N_S = 0.01$, $\kappa = 0.01$, and $\varepsilon = 0.01$. In both figures the red dashed and black dashed lines 
correspond to the maximum values for the TMSV and three-mode Gaussian states.
Thus, we conjecture that $R$ for the TMSV state is larger and smaller than $R$ for the three-mode state when  $N_S \gg N_B$ and $N_S \ll N_B$ respectively.
However, as shown in the following this conjecture is not exactly right.

In order to confirm our conjecture, we define the ratio
\begin{equation}
\label{ratio-1}
r = \frac{R_{max} (\mbox{TMSV})}{R_{max} (3-\mbox{mode})}.
\end{equation}
One can show by straightforward calculation that the asymptotic behaviors of $R_{max} (\mbox{TMSV})$ and $R_{max} (3-\mbox{mode})$ become
\begin{eqnarray}
\label{Rmax_TMSV}
R_{max} ({\mbox{TMSV}}) = \left\{                  \begin{array}{cc}
                                                                        \frac{\kappa N_S}{1 - \kappa} \ln \left( \frac{1 + N_B - \kappa}{N_B} \right) + {\cal O} (1)                            &  \hspace{1.0cm} N_S \gg N_B          \\
                                                                        \frac{\kappa N_S (1 + N_S)}{N_B} \ln \left(1 + \frac{1}{N_S} \right) + {\cal O} \left(\frac{1}{N_B^2} \right)     &   \hspace{1.0cm}  N_B \gg  N_S
                                                                                           \end{array}                                \right.
\end{eqnarray}
and 
\begin{eqnarray}
\label{Rmax_3-mode}
R_{max} ({3-\mbox{mode}}) = \left\{                  \begin{array}{cc}
                                                                        \frac{\kappa N_S}{1 - \kappa + 2 N_B}  + {\cal O} (1)                            &  \hspace{1.0cm} N_S \gg N_B          \\
                                                                        \frac{\kappa N_S }{N_B} \left[ (1 + N_S) \ln \left(\frac{2}{N_S}\right) - N_S \right] + {\cal O} \left(\frac{1}{N_B^2} \right)     &   \hspace{1.0cm}  N_B \gg  N_S
                                                                                           \end{array}                                \right.
\end{eqnarray}
respectively.

\begin{figure}[ht!]
\begin{center}
\includegraphics[height=4.7cm]{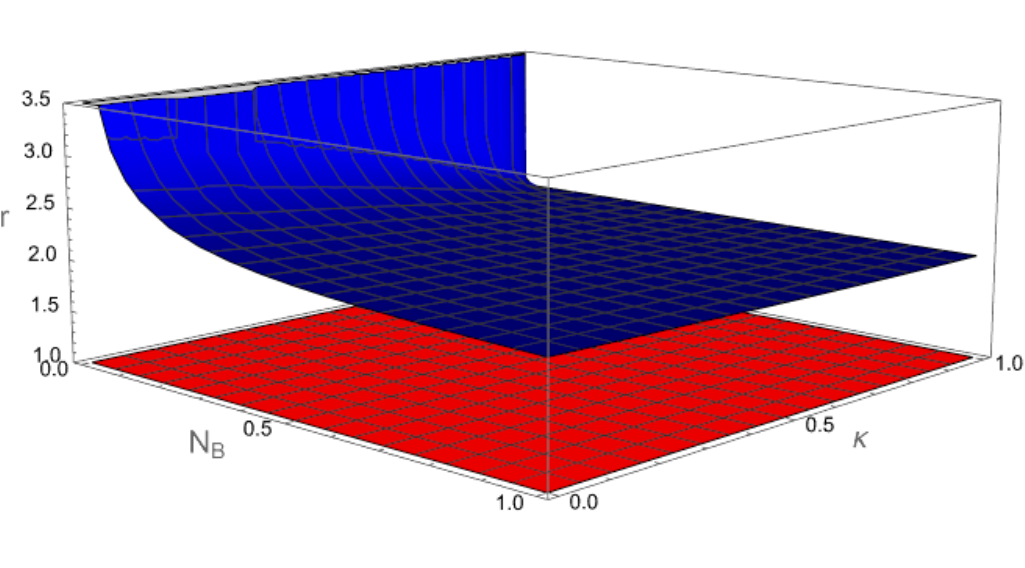} \hspace{.1cm}
\includegraphics[height=5.0cm]{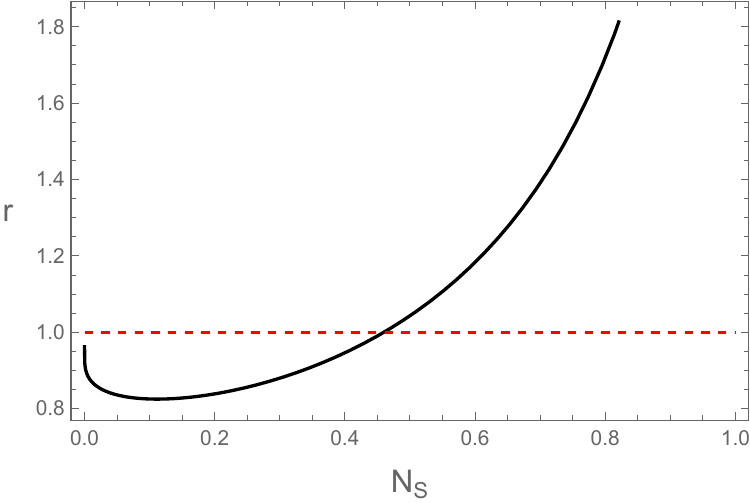}
\caption[fig3]{(Color online) (a) The $N_B$- and $\kappa$-dependence of the ratio $r$ when $N_S \gg N_B$. The red plane corresponds to $r = 1$. This figure shows $r > 1$ in the full range of $N_B$ and $\kappa$ if the condition  $N_S \gg N_B$ holds.
(b) The $N_S$-dependence of the ratio $r$ when $N_B \gg N_S$.   It is shown that $r > 1$ and $r < 1$ when $N_S > N_{S*}$ and $N_S < N_{S*}$, where $N_{S*} = 0.46$.         }
\end{center}
\end{figure}
In Fig. 3 we plot the ratio $r$ when $N_S \gg N_B$ (Fig. 3(a)) and $N_B \gg N_S$ (Fig. 3(b)). Fig. 3(a) shows that $r > 1$ holds in the full-range of $N_B$ and $\kappa$. This means that if $N_S \gg N_B$, $R_{max} (\mbox{TMSV})$ is always larger than $R_{max} (3-\mbox{mode})$.
However, the situation is somewhat different when $N_B \gg N_S$. Fig. 3(b) implies that $R_{max} (\mbox{TMSV})$ is still larger than $R_{max} (3-\mbox{mode})$ if $N_S > N_{S*} = 0.46$ even though $N_S \ll N_B$ holds.
Only when  $N_S < N_{S*}$, $R_{max} (3-\mbox{mode})$ is larger than  $R_{max} (\mbox{TMSV})$. Thus, the error-probability for the TMSV state is less than that for the three-mode Gaussian state in the most region of the parameter space $(N_B, N_S)$.
The reverse situation only occurs when both conditions $N_B \gg N_S$ and $N_S < N_{S*}$ are simultaneously satisfied. In this way the near optimality of TMSV state is still satisfied.

 \section{Conclusion}
In this paper we analyze the AQI with the three-mode Gaussian state (\ref{signal-state}) as an initial state. We also compare the results with those of the TMSV state\cite{asymmetry-1,asymmetry-2}.
It is shown that  the AQI with the three-mode Gaussian state yields more quantum advantage than AQI with TMSV state when both  $N_B \gg N_S$ and $N_S < 0.46$ are simultaneously satisfied. 
If both conditions are not satisfied, the quantum advantage for the TMSV state is more than that for the three-mode Gaussian state. As shown in Ref.\cite{eylee21} similar situation occurs in the SQI. 

What happen when $N$-mode ($N \geq 4$) maximally entangled Gaussian state is used as an initial state in QI? Probably, the region superior to TMSV state may be enlarged in the parameter space $(N_B, N_S)$.
We hope to explore this issue in the future.

{\bf Acknowledgement}:
This work was supported by the National Research Foundation of Korea(NRF) grant funded by the Korea government(MSIT) (No. 2021R1A2C1094580).


\begin{thebibliography}{99}

\bibitem{schrodinger-35} E. Schr\"{o}dinger, {\it Die gegenw\"{a}rtige Situation in der Quantenmechanik}, Naturwissenschaften, 
{\bf 23} (1935) 807.
\bibitem{text} M. A. Nielsen and I. L. Chuang, Quantum Computation and Quantum Information (Cambridge
University Press, Cambridge, England, 2000).
\bibitem{horodecki09} R. Horodecki, P. Horodecki, M. Horodecki, and K. Horodecki, {\it Quantum Entanglement}, Rev. Mod. Phys. 
{\bf 81} (2009) 865 [quant-ph/0702225] and references therein.
\bibitem{teleportation} C. H. Bennett, G. Brassard, C. Crepeau, R. Jozsa, A. Peres and W. K. Wootters, {\it Teleporting
an Unknown Quantum State via Dual Classical and Einstein-Podolsky-Rosen Channles}, Phys.Rev. Lett. {\bf 70} (1993) 1895.
\bibitem{Luo2019}Y. H. Luo et al., {\it Quantum Teleportation in High Dimensions}, Phys. Rev. Lett. {\bf 123} (2019) 070505. [arXiv:1906.09697 (quant-ph)]
\bibitem{superdense} C. H. Bennett and S. J. Wiesner, {\it Communication via one- and two-particle operators on
Einstein-Podolsky-Rosen states}, Phys. Rev. Lett. {\bf 69} (1992) 2881.
\bibitem{clon} V. Scarani, S. Lblisdir, N. Gisin and A. Acin, {\it Quantum cloning}, Rev. Mod. Phys. {\bf 77} (2005)
1225 [quant-ph/0511088] and references therein.
\bibitem{cryptography} A. K. Ekert , {\it Quantum Cryptography Based on Bell’s Theorem}, Phys. Rev. Lett. {\bf 67} (1991)
661.
\bibitem{cryptography2} C. Kollmitzer and M. Pivk, Applied Quantum Cryptography (Springer, Heidelberg, Germany, 2010).
\bibitem{metro17} K. Wang, X. Wang, X. Zhan, Z. Bian, J. Li, B. C. Sanders, and P. Xue, {\it Entanglement-enhanced quantum metrology in a noisy environment}, Phys. Rev. {\bf A97} (2018) 042112. [arXiv:1707.08790 (quant-ph)]
\bibitem{qcreview} T. D. Ladd, F. Jelezko, R. Laflamme, Y. Nakamura, C. Monroe, and J. L. O'Brien, 
{\it Quantum Computers}, Nature, {\bf 464} (2010) 45. [arXiv:1009.2267 (quant-ph)]
\bibitem{computer} G. Vidal, {\it Efficient classical simulation of slightly entangled quantum computations}, Phys. Rev.
Lett. {\bf 91} (2003) 147902. [quant-ph/0301063]
\bibitem{supremacy-1} F. Arute et al.,{Quantum supremacy using a programmable superconducting processor}, Nature {\bf 574} (2019) 505. Its supplementary information is given in arXiv:1910.11333 (quant-ph).
\bibitem{lloyd08} S. Lloyd, {\it Enhanced Sensitivity of Photodetection via Quantum Illumination}, Science, {\bf 321}, 1463 (2008).
\bibitem{hypo1}E. L. Lehmann and J. P. Romano, {\it  Testing Statistical Hypotheses} (Springer, New York, 2008).
\bibitem{hypo2} C. W. Helstrom, {\it Quantum Detection and Estimation Theory, Mathematics in Science and Engineering} (Academic Press, New York, 1976), Vol. 123.
\bibitem{tan08}S.-H. Tan, B. I. Erkmen, V. Giovannetti, S. Guha, S. Lloyd, L. Maccone, S. Pirandola, and J. H. Shapiro, {\it Quantum Illumination with Gaussian States}, Phys. Rev. Lett. {\bf 101}, 253601 (2008). [arXiv:0810.0534 (quant-ph)]
\bibitem{asymmetry-1} M. M. Wilde, M. Tomamichel, S. Lloyd, and M. Berta, {\it Gaussian Hypothesis Testing and Quantum Illumination}, Phys. Rev. Lett. {\bf 119}, 120501 (2017). [arXiv:1608.0699 (quant-ph)]
\bibitem{asymmetry-2} A. Karsa, G. Spedalieri, Q. Zhuang, and S. Pirandola, {\it Quantum illumination with a generic Gaussian source}, Phys. Rev. Research {\bf 2}, 023414 (2020). [arXiv:2005.07733 (quant-ph)]
\bibitem{gaussian1}A. Serafini, {Quantum Continuous Variables} (CRC Press, New York, 2017).
\bibitem{guha09} S. Guha and B. I. Erkmen, {\it Gaussian-state quantum-illumination receivers for target detection}, Phys. Rev. {\bf A 80} (2009) 052310. [arXiv:0911.0950 (quant-ph)]
\bibitem{lopaeva13} E. D. Lopaeva, I. R. Berchera, I. P. Degiovanni, S. Olivares, G. Brida, and M. Genovese, {\it Experimental Realization of Quantum Illumination}, Phys. Rev. Lett. 110 (2013) 153603. [arXiv:1303.4304 (quant-ph)]
\bibitem{barzanjeh15} S. Barzanjeh, S. Guha, C. Weedbrook, D. Vitali, J. H. Shapiro, and S. Pirandola, {\it Microwave Quantum Illumination}, Phys. Rev. Lett. {\bf 114} (2015) 080503. [arXiv:1503.00189 (quant-ph)]
\bibitem{zhang15} Z. Zhang, S. Mouradian, F. N. C. Wong, and J. H. Shapiro, {\it Entanglement-Enhanced Sensing in a Lossy and Noisy Environment}, Phys. Rev. Lett. {\bf 114} (2015) 110506. [arXiv:1411.5969 (quant-ph)]
\bibitem{zhuang17} Q. Zhuang, Z. Zhang, and J. H. Shapiro, {\it Optimum Mixed-State Discrimination for Noisy Entanglement-Enhanced Sensing}, Phys. Rev. Lett. {\bf 118} (2017) 040801. [arXiv:1609.01968 (quant-ph)]
\bibitem{li12} K. Li, {\it Second order asymptotics for quantum hypothesis testing}, Ann. Stat. {\bf 42} (2014) 171. [arXiv:1208.1400 (quant-ph)]
\bibitem{TH12} M. Tomamichel and M. Hayashi, {\it A hierarchy of information quantities for finite block length analysis of quantum tasks},  IEEE Trans. Inf. Theory {\bf 59} (2013) 7693. [arXiv:1208.1478 (quant-ph)]
\bibitem{CMMAB08} J. Calsamiglia, R. M. Tapia, L. Masanes, A. Acin, and E. Bagan, {\it Quantum Chernoff bound as a measure of distinguishability between density matrices: Application to qubit and Gaussian states}, Phys. Rev. {\bf A 77} (2008) 032311. [arXiv:0708.2343 (quant-ph)]\bibitem{JOPS12} V. Jaksic, Y. Ogata, C.-A. Pillet, and R. Seiringer, {\it  Quantum hypothesis testing and non-equilibrium statistical mechanics}, Rev. Math. Phys. {\bf 24} (2012) 1230002. [arXiv:1109.3804 (math-ph)]
\bibitem{DPR15} N. Datta, Y. Pautrat, and C.  Rouz\'{e}, {\it Second-order asymptotics for quantum hypothesis testing in settings beyond i.i.d. - quantum lattice systems and more}, J. Math. Phys. {\bf 57} (2016) 062207. [arXiv:1510.04682 (quant-ph)]
\bibitem{zhang14} S. Zhang, J. Guo, W. Bao, J. Shi, C. Jin, X. Zou, and G. Guo, {\it Quantum illumination with photon-subtracted continuous-variable entanglement}, Phys. Rev. {\bf A 89} (2014) 062309.
\bibitem{fan18} L. Fan and M. S. Zubairy, {\it Quantum illumination using non-Gaussian states generated by photon subtraction and photon addition}, Phys. Rev. {\bf A 98} (2018) 012319.
\bibitem{micro} S. Barzanjeh, S. Guha, C. Weedbrook, D. Vitali, J. H. Shapiro, and S. Pirandola, {\it Microwave Quantum Illumination}, Phys. Rev. Lett. {\bf 114} (2015) 080503. [arXiv:1503.00189 (quant-ph)]
\bibitem{guha09-2} S. Guha, {\it Receiver design to harness quantum illumination advantage}, 2009 IEEE International Symposium on Information Theory, Seoul, Pp. (2009)., pp. 963–967. [arXiv:0902.2932 (quant-ph)]
\bibitem{Jo21-2} Y. Jo, S. Lee, Y. S. Ihn, Z. Kim, and S. Y. Lee, {\it Quantum illumination receiver using double homodyne detection}, Phys. Rev. Research {\bf 3} (2021) 013006. [arXiv:2008.11928 (quant-ph)]
\bibitem{museong23-1} M. Kim, M. -R. Hwang, E. Jung, and D. K. Park, {\it Is entanglement a unique resource in quantum illumination?}, Quant. Inf. Proc. {\bf 22}  (2023) 98. [arXiv:2111.08941 (quant-ph)]
\bibitem{discord1} C. Weedbrook, S. Pirandola, J. Thompson, V. Vedral, and M. Gu, {\it How Discord underlies the Noise Resilience of Quantum Illumination}, New J. Phys. {\bf 18} (2016) 043027. [arXiv:1312.3332 (quant-ph)]
\bibitem{discord2} M. Bradshaw, S. M. Assad, J. Y. Haw, S.-H. Tan, P. K. Lam, and M. Gu, {\it Overarching framework between Gaussian quantum discord and Gaussian quantum illumination}, Phys. Rev. {\bf A 95} (2017) 022333. [arXiv:1611.10020 (quant-ph)]
\bibitem{anti-discord1} M.-H. Yung, F. Meng, X.-M. Zhang, and M.-J. Zhao, {\it One-Shot Detection Limits of Quantum Illumination with Discrete Signals}, npj Quantum Inf. {\bf 6} (2020) 75. [arXiv:1801.07591 (quant-ph)]
\bibitem{Jo21-1} Y. Jo, T. Jeong, J. Kim, D. Y. Kim, Y. S. Ihn, Z. Kim, and S. Y. Lee, {\it Quantum illumination with asymmetrically squeezed two-mode light}, arXiv:2103.17006 (quant-ph).
\bibitem{palma18}G. D. Palma and J. Borregaard, {\it minimum error probability of quantum illumination}, Phys. Rev. {\bf A 98} (2018) 012101. [arXiv:1802.02158 (quant-ph)]
\bibitem{nair20} R. Nair and M. Gu, {\it Fundamental limits of quantum illumination}, Optica {\bf 7} (2020) 771. [arXiv:2002.12252 (quant-ph)]
\bibitem{brad21} M. Bradshaw, L. O. Conlon, S. Tserkis, M. Gu, P. K. Lam, and S. M. Assad, {\it Optimal probes for continuous variable quantum illumination}, Phys. Rev. {\bf A 103} (2021) 062413. [arXiv:2010.09156 (quant-ph)] 
\bibitem{eylee21} E. Jung and D. K. Park, {\it Quantum Illumination with three-mode Gaussian State},  Quant. Inf. Proc. {\bf 21}  (2022) 71. [arXiv:2107.05203 (quant-ph)]
























\end{thebibliography}
\end{document}